\documentclass[aps,prl,twocolumn,superscriptaddress,showpacs]{revtex4-1}

\usepackage{graphicx}
\usepackage{amsmath}
\usepackage{amsfonts}
\usepackage{bm}

\usepackage{times}

\usepackage{color}

\newcommand{\ave}[1]{\langle #1 \rangle}

\begin{document}
\title{Vibrational Andreev bound states in magnetic molecules}

\author{Denis Gole\v{z}}

\affiliation{Jo\v{z}ef Stefan Institute, Jamova 39, SI-1000 Ljubljana, Slovenia}

\author{Janez Bon\v{c}a}

\affiliation{Jo\v{z}ef Stefan Institute, Jamova 39, SI-1000 Ljubljana, Slovenia}
\affiliation{Faculty  of Mathematics and Physics, University of Ljubljana, Jadranska 19, SI-1000 Ljubljana, Slovenia}

\author{Rok \v{Z}itko}

\affiliation{Jo\v{z}ef Stefan Institute, Jamova 39, SI-1000 Ljubljana, Slovenia}
\affiliation{Faculty  of Mathematics and Physics, University of Ljubljana, Jadranska 19, SI-1000 Ljubljana, Slovenia}
	
\begin{abstract}
We predict the existence of vibrational Andreev bound states in
deformable magnetic molecules on superconducting surfaces. We discuss
the Anderson impurity model with electron-phonon coupling to a
realistic anharmonic vibrational mode that modulates the tunneling
barrier and show that the vibronic features are spectroscopically most
visibile near the transition point between the Kondo-screened singlet
and the unscreened doublet ground state. We find competing tendencies
between phonon hardening due to anharmonicity and softening due to
coupling to electrons, contrary to the Anderson-Holstein model and
other models with harmonic local phonon mode where the vibrational
mode is always softened. In addition, we find that the singlet and
doublet many-body states may experience very different effective
phonon potentials.
\end{abstract}

\pacs{72.15.Qm, 73.20.Hb, 73.40.Gk, 75.75.-c, 74.55.+v}

\maketitle

Molecules that conduct electrical current \cite{bumm1996,nitzan2001,
song2011, zimbovskaya2011},
when embedded in a junction
between two metal electrodes \cite{reed1997}, can become the
active element in a circuit, such as a rectifier
\cite{aviram1974,Diez-Perez2009} or a memory element
\cite{chen1999onoff}. Alternatively, molecules  
deposited on a metal substrate can be probed by a scanning
tunneling microscope to study their diffusion \cite{schunack2002prl},
conformation changes \cite{moresco2001b,donhauser2001}, dissociation
\cite{stipe1997}, and chemical reactions \cite{wo1999, hla2000}.
Strong coupling between electronic and vibrational degrees of freedom
plays a critical role for the molecule's
functional properties \cite{galperin2006,
leturcq2009}. The electron-phonon (e-ph) coupling renormalizes the
electron-electron (e-e) interaction \cite{cornaglia2004,cornaglia2007}
and, if large enough, leads to an effective attractive interaction
\cite{cornaglia2004,cornaglia2005,mravlje2005}. Vibrational modes are detected in the differential
conductance spectra as spectral features at characteristic
frequencies \cite{jaklevic1966, park2000, lorente2000,yu2004c},
that serve as ``molecular fingerprints''
\cite{stipe1998,stipe1999,ho2002}. In magnetic molecules on normal
metal surfaces, the
low-temperature spectra exhibit zero-bias anomalies due to the Kondo
screening of the local moment \cite{zhao2005,scott2010}, while magnetic
molecules adsorbed on superconductors exhibit sharp
spectral peaks inside the gap (Andreev bound states,
ABS) due to the competition between the Kondo effect and the
electron pairing
\cite{satori1992,sakai1993,yoshioka2000,oguri2004,Bauer2007,hecht2008},
as has been recently experimentally demonstrated \cite{franke2011}.
Since molecules are deformable, the vibrational modes need to be
taken into account for a comprehensive description of all features
that may occur in the subgap part of the spectrum.

In this work we study a realistic model of a deformable magnetic
molecule in contact with a superconductor. The vibrational mode is
described using the anharmonic Morse potential
\cite{morse1929,koch2005} and the displacement exponentially modulates
the tunneling barrier \cite{mravlje2008}. The anharmonicity is
required to remove unphysical infinite-displacement solution found in
the harmonic approximation, while the exponential modulation removes
the fluctuating-sign problem of the lowest-order linear coupling; both
choices are also closer to reality. We focus on the case of a weakly
bound adsorbate with external (center-of-mass) vibrational mode whose
energy is comparable with the superconducting gap, so that vibronic
features occur inside the gap. In particular, we show that the
molecule spectral function features vibrational side-peaks in addition
to the main ABS peaks. The side-peaks are visible for generic model
parameters even for moderate, experimentally relevant e-ph coupling
strength. Because the peak width is due only to thermal broadening and
experimental noise, this setup permits very precise determination of
the phonon frequency renormalization due to the e-ph coupling and the
e-e interactions.

We describe the system with the impurity model
$H=H_{\mathrm{band}}+H_{\mathrm{mol}}+H_\mathrm{osc}+H_{\mathrm{coup}}$.
Here $H_{\mathrm{band}}=\sum_{k\sigma}
\epsilon_{k}c_{k,\sigma}^{\dagger}c_{k,\sigma} + \Delta
\sum_{k,\sigma}
(c_{k,\sigma}^{\dagger}c_{-k,-\sigma}^{\dagger}+\text{h.c.})$, where
$c_{k,\sigma}$ are the conduction-band electron operators with
momentum $k$, spin $\sigma$ and energy $\epsilon_k$, and $\Delta$ is the superconducting
gap. $H_{\mathrm{mol}}=\epsilon (n_\uparrow+n_\downarrow)+U
n_{\uparrow}n_{\downarrow}$ is the molecule Hamiltonian, where
$n_{\sigma}=d_{\sigma}^{\dagger}d_{\sigma}$, $\epsilon$ is the on-site
energy, and $U$ the e-e repulsion. The quantity
$\delta=\epsilon+U/2$ measures the deviation of the system from the
particle-hole (p-h) symmetric point. The displaced molecule feels 
a realistic Morse potential of the form
\begin{equation}
V_{\mathrm{Morse}} = D_{e} \left[ 1-\exp\left(-b \hat{x}\right) \right]^{2},
\end{equation}
where $D_{e}$ is the well depth and $b$ controls its width. The
oscillator Hamiltonian is thus $H_\mathrm{osc}={\hat
p}^2/2m + V_{\mathrm{Morse}}$. We define the harmonic frequency as
$\omega_{0}=b \sqrt{2 D_{e}}/m$, and the displacement and momentum
operators as $\hat{x}=(\hat{a}+\hat{a}^{\dagger}) \sqrt{1/2m\omega_{0}}$
and $\hat{p}=i(\hat{a}^\dagger -\hat{a}) \sqrt{m\omega_0/2}$,
where $\hat{a}$ and $\hat{a}^{\dagger}$ are the phonon ladder operators. 
The shape of the potential is then fully
described by two parameters, $D_{e}$ and $\omega_0$. The harmonic
potential is recovered in the limit $D_{e}\rightarrow\infty$, keeping
$\omega_0$ constant.  In the coupling part
$H_{\mathrm{coup}}=V(x)
\sum_{k,\sigma}(c_{k\sigma}^{\dagger}d_{\sigma}+\mathrm{h.c.})$, the
tunneling term is exponentially modulated by the molecular
displacement:
$$V(x)=V_{0} \exp \left[ -g
(\hat{a}+\hat{a}^{\dagger}) \right],$$
where $g>0$ is the e-ph coupling constant. The hybridization strength
at zero displacement is characterized by $\Gamma_0=\pi \rho V_0^2$,
where $\rho$ is the density of states in the band in the normal state.
In all numerical calculations presented in this work, we set $U=1$,
$\Delta=0.04$, and $\omega_0=0.01$.

In the absence of e-ph coupling ($g=0$) and close to the particle-hole
symmetric point ($\delta\sim 0$), the ground state of the system
depends on the relative values of the Kondo temperature $T_K$ and the
BCS gap $\Delta$ \cite{satori1992,sakai1993,yoshioka2000}. In the
limit $T_{K}\gg\Delta$, the local moment is screened by non-paired
conducting electrons. The ground state is then a spin singlet (S)
many-body Kondo state \cite{satori1992}. In the opposite limit of
$T_{K}\ll\Delta$, the formation of the Cooper pairs leaves no
low-energy electrons available to screen the impurity spin. The ground
state is then a spin doublet (D) unscreened many-body state. For $T_K
\sim \Delta$, there is a level crossing between the two different
ground states. The subgap part of the spectral function for $g=0$ has
generically two peaks, located symmetrically with respect to the Fermi
level (here fixed at $\omega=0$). These features, often referred to as
the Andreev bound states or Shiba states, correspond to the
transitions between the S and D many-body states. These states are
``bound'' in the sense that they correspond to excitations below the
continuum of quasiparticle states and that the corresponding spectral
weight is localized around the impurity site. The transitions can
occur either by injecting an electron ($\omega>0$ peak) or by removing
it ($\omega<0$ peak).  Away from the p-h symmetric point, the weights
of these two peaks are different \cite{hecht2008}. Their spectral
weights go to zero as the S-D energy difference tends toward $\Delta$
in both small-$\Gamma$ ($T_K \ll \Delta$) and large-$\Gamma$ ($T_K \gg
\Delta$) hybridization limits and is maximal near the S-D level
crossing \cite{Bauer2007}, which is signaled by the crossing of the
subgap peaks in the spectrum \cite{yoshioka2000,franke2011}.

We now consider the effect of the e-ph coupling ($g>0$) on the
subgap states. For small coupling the hybridization is renormalized
as $\Gamma_0 \to \tilde \Gamma=\Gamma(g, \ave{x})$, where $\ave{x}$ is
the expectation value of the displacement that is negative (the molecule approaches the surface) and linear
in $g$ [see Fig.{3}(g)-(i)], thus the enhancement of the hybridization
is approximately quadratic. For large coupling this mean-field
approximation is no longer accurate, but the trend toward stronger
effective $\Gamma$ remains. 
In addition, if the phonon
frequency is smaller that the BCS gap, $\omega_{0}<\Delta$,
entirely new features (vibronic ABS sidepeaks) appear in the subgap
part of the molecule spectral function, as we show in the following.

The full problem is solved using the numerical renormalization group
\cite{wilson1975,bulla2008} with extensions for superconducting
bands \cite{satori1992,sakai1993,yoshioka2000}. This technique
treats all interactions (e-ph coupling, e-e repulsion, BCS pairing) on
equal footing. The calculations have been performed for the
discretization parameter $\Lambda=4$ and are fully converged with
respect to the phonon cutoff and the number of states kept in the
truncation.

\begin{figure}
\centering
\includegraphics[width=8cm,clip]{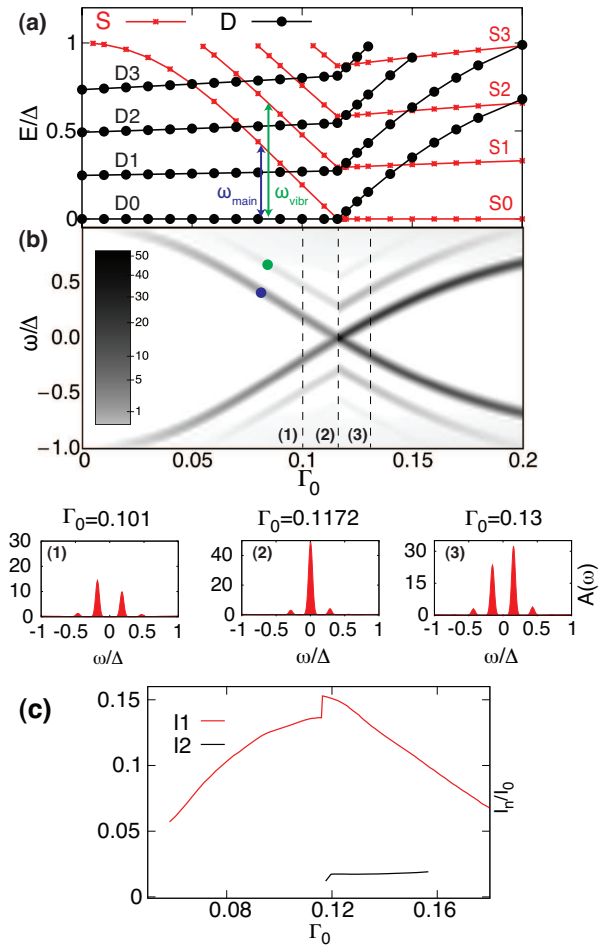}
\caption{(Color online) (a) Energy of the subgap singlet (S) and
doublet (D) states as a function of the hybridization strength
$\Gamma_0$.
(b) Subgap part of the spectral function.
Subfigures show cross-sections at particular values of $\Gamma_0$,
indicated by dashed lines.
(c) Visibility of the first and second vibronic side-peaks.
Parameters are $g=0.05$, $D_e=0.5$, and $\delta=0.1$.
}
\label{Fig:gamma_sweep}
\end{figure} 

\begin{figure}
 \centering
\includegraphics[width=8cm,clip]{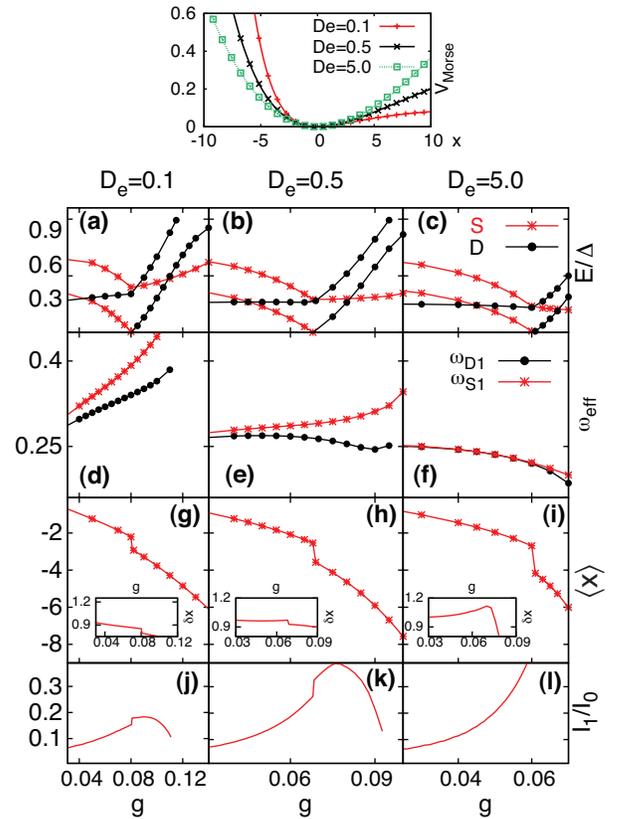}
 \caption{(Color online) Top panel: bare (non-renormalized) Morse
 potential for different parameter of the anharmonicity.
 Bottom panel:
 (a,b,c) Energy of the subgap singlet (S) and
 doublet (D) states versus electron-phonon coupling $g$ for different
 parameter of the anharmonicity.
 Parameters: $\Gamma_0=0.1$, $\delta=0$.
(d,e,f) Effective phonon frequencies of D and S states.
 (g,h,i) Displacement and fluctuations (insets) of the displacement.
 (j,k,l) Visibility of first vibronic side-peak.}
\label{Fig:gg_com}
 \end{figure}

We first consider the evolution of the subgap states as a function of
the hybridization strength $\Gamma_0$, keeping the e-ph coupling $g$
constant, see Fig.~\ref{Fig:gamma_sweep}(a). This is motivated
by the experimental realization of the singlet-doublet crossing,
which is tuned by small changes in the molecule-substrate coupling \cite{franke2011}.
The calculation has been performed for generic parameter values, in particular away from the
p-h symmetric point. The level crossing between the doublet and singlet ground
states D0 and S0 occurs at $\Gamma_0=\Gamma_{0c}$. All other excited subgap
states, one series of D states (D1, D2, {\ldots}) and another of S
states (S1, S2, \ldots), are phonon
induced. For a harmonic potential and very weak e-ph coupling, these
vibronic states would form ladders with equidistant spacing
$\omega_0$. For the anharmonic Morse potential, however, the exact
solution for the vibrational energy levels includes a quadratic
correction term \cite{morse1929} and, furthermore, the effective
energy spacing is renormalized by the e-ph coupling \cite{Hewson2002,
meyer2002, jeon2003, mravlje2008}, see also Fig.~\ref{Fig:gg_com}(d)-(f).

The subgap part of the spectral function,
Fig.~\ref{Fig:gamma_sweep}(b), demonstrates that in addition to the
main spectral peak (S0-D0 transitions),
other phonon-induced side-peaks (S0-D1, D0-S1, etc. transitions) are
also present and have sufficient spectral weight to be observable.
The subfigures Fig.~\ref{Fig:gamma_sweep}(b1-b3) show spectral curves
at three characteristic parameter regimes; first side peaks are clearly
resolved in all cases. The ratio between the spectral weight of the
side peaks and that of the main ABS peaks ({\it i.e.}, the visibility),
Fig.~\ref{Fig:gamma_sweep}(c), is maximal near the level crossing. 
We also find that the visibility is largest at the p-h symmetric point
($\delta=0$) and is reduced somewhat away from it (results not shown).

\begin{figure}
 \centering 
\includegraphics[width=8cm,clip]{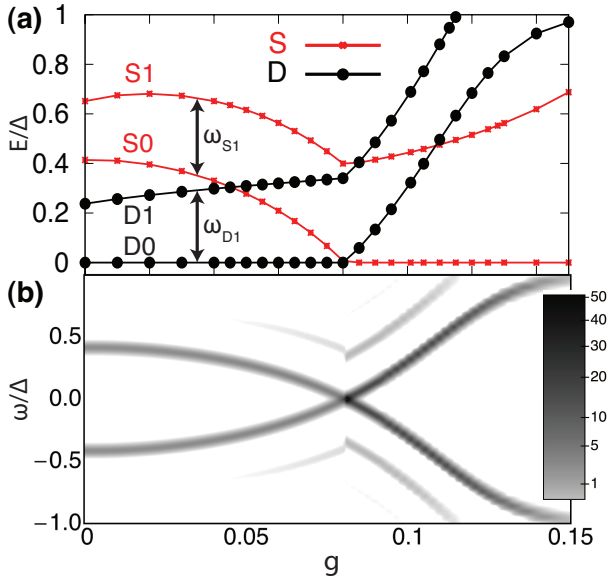}
 \caption{(Color online) (a) Energy of the subgap singlet (S) and
 doublet (D) states as a function of the electron-phonon coupling $g$.
 (b) Subgap part of the spectral function. Parameters are 
 $\Gamma_0=0.1$, $D_{e}=0.1$, and $\delta=0$.
 }
 \label{Fig:gg_sweep}
 \end{figure}

Alternatively, the S-D transition can be induced by increasing the
e-ph coupling $g$ at fixed $\Gamma_0$, see Fig.~\ref{Fig:gg_sweep}.
The vibrational excited states exhibit an unexpected feature: with
increasing e-ph coupling, the effective phonon frequency increases
(the phonon mode hardens). The renormalization of the phonon frequency
has been noted in previous studies of impurity models with vibration
modes \cite{Hewson2002,meyer2002,jeon2003,mravlje2008}, where the
phonon mode softens. This is the case both for the coupling to charge
(Anderson-Holstein model) and for the coupling to the center-of-mass
modes. 
The phonon hardening is a characteristic feature of the anharmonic
potential and results from the displacement of the oscillator to the
part of the potential with higher second derivative, while the
softening results from the electron tunneling fluctuations
\cite{balseiro2006} and occurs generically. Both tendencies are
present in our model and the dominant effect depends on the
value of the parameter $D_e$.

In order to compare the degree of the anharmonicity, the bare
(non-renormalized) potential profiles are shown in the top panel of
Fig.~\ref{Fig:gg_com}. In Fig.~\ref{Fig:gg_com}(a)-(c) the energy
dependence of the Andreev states shows the S-D transition as a
function of the e-ph coupling $g$ for different values of $D_{e}$. The
renormalized phonon frequencies, Fig.~\ref{Fig:gg_com}(d)-(f), indicate
that for $D_e=0.1$, the potential is strongly anharmonic and the
phonon mode hardens, while for $D_e=5$, in the harmonic limit, it
softens. We also observe remarkably large differences in the D and S
sectors, which are the most pronounced for intermediate anharmonicity
where, as a function of $g$, the phonon mode softens in the D sector,
while it hardens in the S sector [see Fig.~\ref{Fig:gg_com}(e)].
The deformation of the molecules as a function of $g$ is shown in
Fig.~\ref{Fig:gg_com}(g)-(i) [notice that the horizontal ranges are
different].  In the harmonic limit the e-ph coupling leads to
unphysically strong
deformation already at small values of $g$, while
for anharmonic potential, the displacement of the molecule
is constrained by the repulsive part of the Morse potential.
The insets represent the fluctuations of the displacement, $\delta x =
\left( \ave{x^2}-\ave{x}^2 \right)^{1/2}$, which in the harmonic limit
strongly increase at the transition and then rapidly drop at higher
$g$, while for strongly anharmonic potential they decrease
monotonously and there is no enhancement at the transition.
Independent of the anharmonicity parameter $D_{e}$, the visibility
grows with increasing e-ph coupling for $g<g_{c}$, see
Fig.~\ref{Fig:gg_com}(j)-(l). For larger e-ph coupling the visibility
starts to decrease, since the energy difference between the ground
state and the side-peak states approaches the BCS gap
\cite{Bauer2007}. The maximum of the visibility thus generically
occurs near the S-D transition.
The visibility is larger for high $D_e$ values, i.e., in the harmonic
limit.

\begin{figure}
\includegraphics[width=8cm,clip]{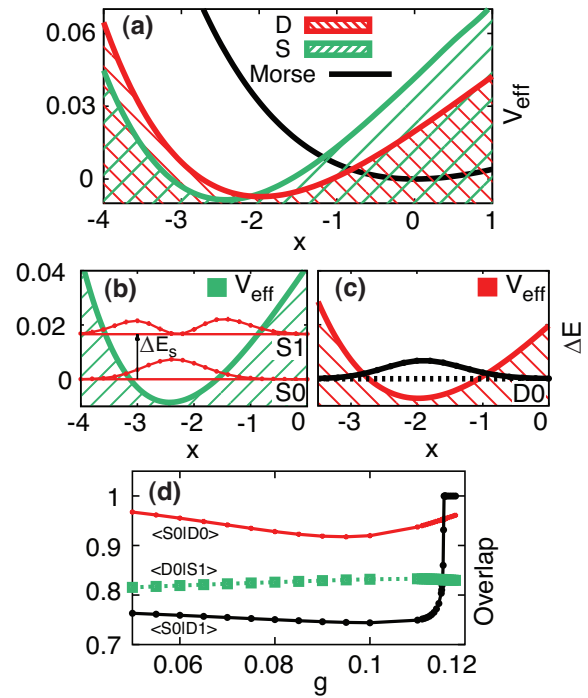}
\caption{(Color online) (a) Effective (renormalized) potential of S
and D subgap states and bare (non-renormalized) Morse potential. (b,c)
Oscillator distribution $\rho(x)$ for ABS in the effective potential for S and
D states. The probability is not normalized. (d) Overlap of the phonon
part of the wavefunction between different subgap states. Parameters
are $\Gamma_0=0.1$, $D_e=0.1$, $\delta=0$.} \label{Fig:overlap}
\end{figure}

The oscillator distribution functions $\rho(x)$ and the effective
potentials $V_{\mathrm{eff}}(x)$,
calculated using the reduced phonon density matrix \cite{hewson2010},
are shown in Fig.~\ref{Fig:overlap}.
The minimum of the effective potential is significantly shifted due to
the e-ph coupling to its new equilibrium position and
we find that the effective potential is not the same for S and D
states. This implies that the oscillator parameters are renormalized
differently in the two spin sectors due to different
electron tunneling rates in the unscreened D and Kondo screened S
many-body states.
In the weak coupling (small-$g$) regime the effect
of the phonons on the subgap part of the spectral function can be
understood using the Frank-Condon principle \cite{Condon1928}, which
states that the transition between the vibrational states is more
likely to happen if the vibrational wave functions overlap more
significantly. The overlap of the effective vibronic part of the wave
functions, $\phi(x)=\sqrt{\rho(x)}$, is proportional to the transition
probability between the states. For the lowest-lying singlet $S0$ and
doublet $D0$ states, and the excited singlet $S1$ and doublet $D1$
states, the overlaps are represented in Fig.~\ref{Fig:overlap}(d).
Within the Franck-Condon approximation the overlap ratio $|\langle S0 |
D1 \rangle|^2/|\langle S0 | D0 \rangle|^2$ is proportional to the visibility
and provides good agreement with the actual values,
Fig.~\ref{Fig:gg_com}(j), for small electron-phonon couplings $g$.
However, since the Franck-Condon principle is based on the
Born-Oppenheimer approximation, polaronic effects are neglected and,
therefore, it cannot explain the features in the spectral function for
e-ph interaction close to or beyond the level crossing. The sudden
increase in the overlap between the S0 and D1 states for $g\approx
0.115$, coincides with the transition of the D1 state into the
continuum, where the excited D1 state strongly mixes with the direct
product states consisting of the S0 state and an additional continuum
low-energy quasiparticle, thus its vibrational properties are
essentially those of the S0 state, which explains the perfect overlap.

We have shown that due to the deformability of the molecule additional
vibronic states occur inside the BCS gap, which are clearly visible in
the subgap part of the spectral function. We find that generically the
intensity of the side peak reaches a maximum close to the ground-state
level crossing, but that the peak intensity depends on the
system parameters: it increases near the particle-hole symmetric point,
for large $D_e$, and for large e-ph coupling $g$.
The increase in the visibility for weak e-ph 
coupling can be described using the Franck-Condon principle.

To test the predictions of this work, we propose as candidate systems
planar macrocyclic molecules which form coordination complexes with weak metal-ligand bonds, so that the magnetic ions support low frequency "rattling" modes, while appropriate surfaces are elemental BCS superconductors.

\begin{acknowledgments}
The authors acknowledge discussions with Jernej Mravlje, Toma\v{z}
Rejec, Anton Ram\v{s}ak and Peter Prelov\v{s}ek, and the support from
the Slovenian Research Agency (ARRS) under Program P1-0044.
\end{acknowledgments}

\bibliography{BibTex/NRG.bib}
\end{document}